\documentstyle[aps,multicol,prb,epsf]{revtex} 
\begin{document} 
\title{Giant magnetothermopower of magnon-assisted transport in
ferromagnetic tunnel junctions} 
\author{Edward McCann and Vladimir I. Fal'ko} 
\address{Department of Physics, Lancaster University,
Lancaster, LA1 4YB, United Kingdom
\\and\\
Grenoble High Magnetic Field Laboratory,
25 Avenue des Martyrs, F-38042 Grenoble Cedex 9, France}
\date{\today} 
\maketitle
\begin{abstract}
{We present a theoretical description of the thermopower due to
magnon-assisted tunneling in a mesoscopic tunnel junction
between two ferromagnetic metals.
The thermopower is generated in the course of thermal
equilibration between two baths of magnons, mediated by electrons.
For a junction between two ferromagnets with
antiparallel polarizations, the ability of magnon-assisted tunneling
to create thermopower $S_{AP}$
depends on the difference between the size $\Pi_{\uparrow , \downarrow}$
of the majority and minority
band Fermi surfaces and it is proportional to a temperature dependent
factor $(k_{B}T/\omega _{D})^{3/2}$ where $\omega _{D}$ is the
magnon Debye energy.
The latter factor reflects the fractional change in the net
magnetization of the reservoirs due to thermal magnons
at temperature $T$ (Bloch's $T^{3/2}$ law).
In contrast, the contribution of magnon-assisted tunneling to
the thermopower $S_P$ of a junction with parallel polarizations is negligible.
As the relative polarizations of ferromagnetic layers can be manipulated by
an external magnetic field, a large difference
$\Delta S = S_{AP} - S_P \approx S_{AP}
\sim - (k_B/e) f\;\!\!(\Pi_{\uparrow},\Pi_{\downarrow})
(k_BT/\omega _{D})^{3/2}$
results in a magnetothermopower effect.
This magnetothermopower effect becomes giant in the extreme case of
a junction between two half-metallic ferromagnets,
$\Delta S \sim - k_B/e$.
}\end{abstract}

 
\begin{multicols}{2}
\bibliographystyle{simpl1}

\section{Introduction}

Spin polarized transport has recently been the subject of
intense theoretical and experimental interest.\cite{pri95}
The mismatch of spin currents at the interface between two
ferromagnetic (F) electrodes with antiparallel spin polarizations
produces a larger contact resistance than a junction with parallel
polarizations, leading to tunneling magnetoresistance
in F-F junctions\cite{jul75,moo00}
and giant magnetoresistance (GMR) in multilayer structures.\cite{bai88,pra91}
Systems displaying GMR have shown other magnetotransport effects
including substantial magnetothermopower
\cite{sak91,con91,pir92,avdi93,shi93,pir93,shi93a,shi96,sat98,bai2000}
with a strong temperature dependence.
Thermoelectric effects have also been discussed in the context of
spin injection across a ferromagnetic-paramagnetic junction.\cite{j+s87}

The Mott formula\cite{zim64}
$S=- \;\!\! (\pi^{2}k_{B}^{2}T/3e) 
( \partial \:\!\! \ln \:\!\! \sigma (\epsilon )/
\partial \epsilon )_{\epsilon _{F}}$
relates the thermopower $S$ of a system to the derivative
with respect to energy of the electrical conductivity,
$\sigma (\epsilon )$, near the Fermi energy, $\epsilon_{F}$,
so that, in metals, $S$ typically
contains a small parameter such as $k_{B}T/\epsilon_{F}$.
In magnetic multilayers with highly transparent
interfaces, the Mott formula has been used as a basis
for theories of transport that explain the origin of the
magnetothermopower effect as due to either the
difference in the energy dependence of the density of states for
majority and minority spin bands in ferromagnetic layers,\cite{shi96,tsy99}
or a different efficiency of electron-magnon scattering for carriers in
opposite spin states.\cite{pir92}
In particular, the electron-magnon interaction in a ferromagnetic layer
was incorporated to explain
the observation \cite{pir92} of a
strong temperature dependence of $S(T)$ and gave, theoretically, a
much larger thermopower in the parallel configuration of multilayers with
highly transparent interfaces than in the antiparallel one, $S_{P}\gg S_{AP}$.
For tunnel junctions, magnon-assisted processes have been studied both
theoretically\cite{bra98} and experimentally\cite{tsu71}
with a view to relate nonlinear $I(V)$ characteristics to the density of
states of magnons $\Omega (\omega)$ as $d^2I/dV^2 \propto \Omega (eV)$. 

In this paper we investigate a model of the electron-magnon interaction
assisted thermopower in a mesoscopic size ferromagnet/insulator/ferromagnet
tunnel junction, which yields a different prediction.
The bottle-neck of both charge and heat transport lies in a
small-area tunnel contact between ferromagnetic metals held at different
temperatures.
The thermopower is generated in the course of thermal
equilibration between two baths of magnons as mediated by electrons,
and, in the relatively high resistance antiparallel (AP) configuration of
a ferromagnetic tunnel junction,
it depends on the
difference between the size of the majority and minority band Fermi surfaces.
For a momentum-conserving tunneling model we find that
\begin{eqnarray}
S_{AP} &\approx& - \frac{k_B}{e}
\frac{\left( \Pi_{+} - \Pi_{-} \right)}{2\Pi_{-}}
\, \delta m(T) ,
\label{Sintro}
\end{eqnarray}
\begin{eqnarray}
\delta m(T) = \frac{3.47}{\xi} 
\left( \frac{k_BT}{\omega_D} \right)^{3/2} ,
\label{mt0}
\end{eqnarray}
where $\Pi_{+}$($\Pi_{-}$) is the area of the maximal cross-section
of the Fermi surface of majority (minority) electrons in the plane parallel
to the interface ($\Pi_{+} > \Pi_{-}$).
The function $\delta m(T)$ is the fractional change in the
magnetization of the reservoirs due to thermal magnons at temperature $T$
(Bloch's $T^{3/2}$ law),
$\xi$ is the spin of localized moments,
and $\omega_D$ is the magnon Debye energy.
On the other hand, we find that the contribution of magnon-assisted tunneling
to the thermopower of a parallel configuration is negligible.

As an extreme example, the magnetothermopower effect is most pronounced
in the case of half-metallic ferromagnets,
where the exchange spin splitting $\Delta$
between the majority and minority conduction bands is greater
than the Fermi energy $\epsilon _{F}$ measured from the bottom of the
majority band, and the Fermi density of states in the minority band is zero.
In the antiparallel configuration of such a junction, where the
emission/absorption of a magnon would lift the spin-blockade of electronic
transfer between ferromagnetic metals, we predict a large thermopower
effect, whereas in the lower-resistance parallel configuration thermopower
is relatively weak:\cite{mcc02}
\begin{equation}
S_{AP}\approx -0.64\frac{k_{B}}{e}\, ; \qquad
\frac{S_{P}}{S_{AP}}\sim \frac{k_{B}T}{\epsilon_{F}}.
\label{MainResHM}
\end{equation}
This is because the contribution of magnon-assisted transport to the
thermopower in the parallel configuration $S_P$ is zero and the
thermopower only arises from the energy dependence of the electronic
density of states near the Fermi energy.

The magnon-assisted processes that we consider are similar to those
discussed in Refs.~\onlinecite{bra98} in
relation to the magnon contribution to the nonlinear conductance.
In a ferromagnetic tunnel junction in the antiparallel configuration,
the elastic contribution to the conductance is suppressed by the mismatch of
spin currents at the interface.
However it is possible to lift spin current blockade while conserving the
overall spin of the system by emitting or absorbing magnons.
For example, the change of spin that occurs when a minority carrier flips
and occupies a majority state is compensated by an opposite
change of spin due to magnon emission.
As a result, the spin current carried by electrons crossing an interface
between oppositely polarized ferromagnets is carried further by the
flow of magnons (spin waves).

Microscopically, a typical magnon-assisted process that contributes to
the thermopower in the antiparallel configuration,
Eqs.~(\ref{Sintro}) and (\ref{MainResHM}), is shown schematically in Figure~1.
Here the majority electrons on the left hand side of the
junction are `spin-up' and the majority electrons on the right are `spin-down'.
The transition begins with a spin-up majority electron on the left,
that then tunnels through the barrier (without spin flip) into an
intermediate, virtual state with spin-up minority polarization on the right
(Figure~1(a)).
In the final step, Figure~1(b), the electron emits a magnon
and incorporates itself into a previously unoccupied state in the
spin-down majority band on the right.
In our approach, we take into account such inelastic tunneling processes
that involve magnon emission and absorption on both sides of the
interface, as well as elastic electron transfer
processes, in order to obtain a balance equation 
for the current $I(V,\Delta T)$ as a function of bias voltage, $V$, and of the
temperature drop, $\Delta T$.
In the linear response regime the electrical current may be written as
\begin{eqnarray}
I = G_V V + G_{\Delta T} \Delta T ,
\label{gvgt}
\end{eqnarray}
where $G_V$ is the electrical conductance and $G_{\Delta T}$
is a thermoelectric coefficient describing the response to a
temperature difference.
Under conditions of zero net current, the thermopower coefficient is
\begin{eqnarray}
S = - \frac{V}{\Delta T} = \frac{G_{\Delta T}}{G_V} .
\label{defS}
\end{eqnarray}
%

%
\begin{figure}
\hspace{0.05\hsize}
\epsfxsize=0.8\hsize
\epsffile{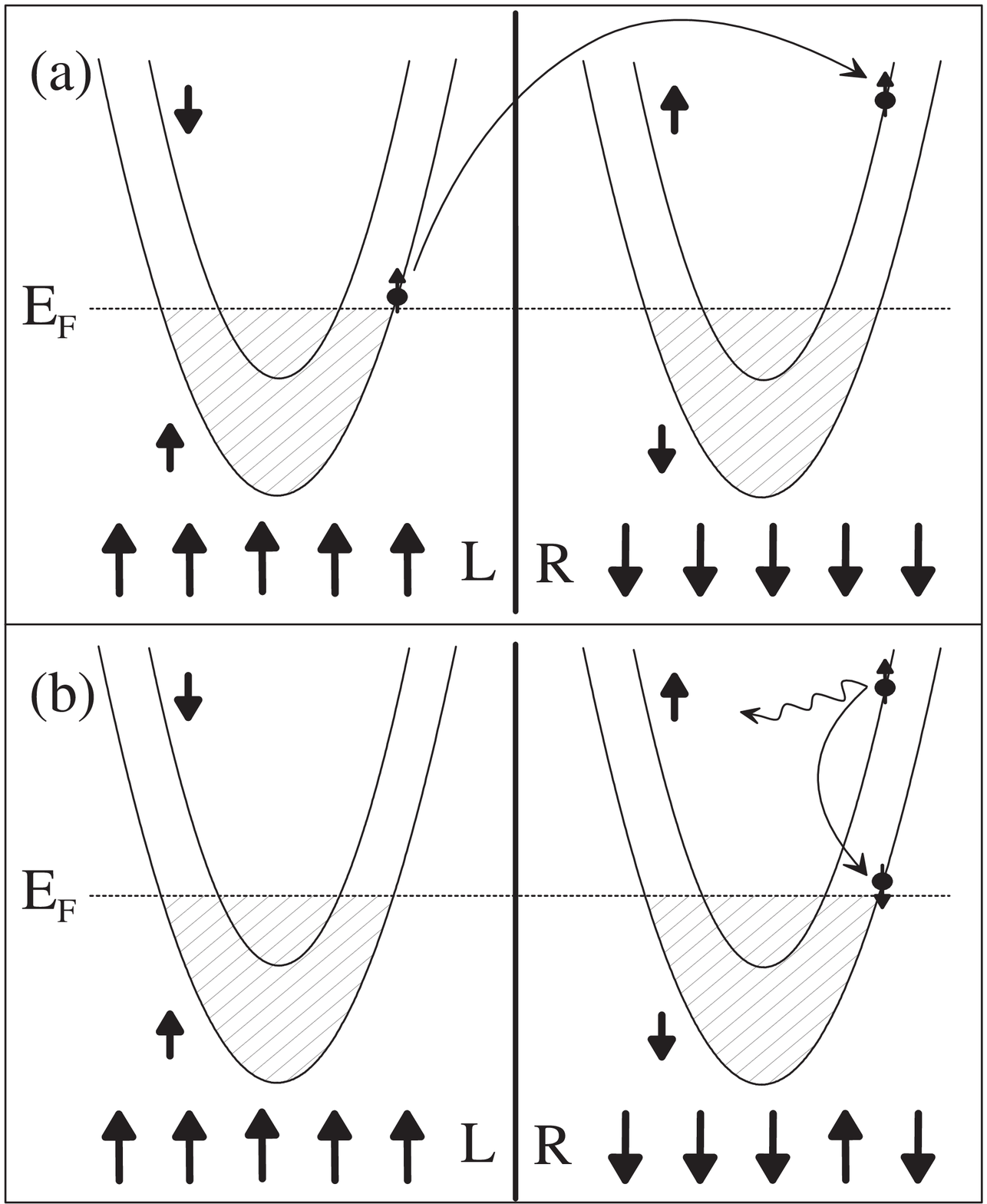}
\refstepcounter{figure}
\label{figure:1}

{\setlength{\baselineskip}{10pt} FIG.\ 1.
Schematic of magnon-assisted tunneling via an intermediate minority state.
This example shows a transition from an initial majority state on the left to
a final majority state on the right
across a junction in the anti-parallel configuration.
(a) The process begins with a
spin-up majority electron on the left, that then tunnels
through the barrier (without spin flip) into an intermediate, virtual state
with spin-up minority polarization on the right.
(b) The electron emits a magnon (wavy line) and incorporates itself into
a previously unoccupied state in the spin-down majority band on the right.
}
\end{figure}
%

The paper is organised as follows.
In Section~\ref{SECTanti} we introduce the model and technique used for
describing a tunnel junction and we calculate the thermopower
in the antiparallel (AP) configuration.
We present a detailed description of two different models of the interface:
a uniformly transparent interface where the component of momentum
parallel to the interface
is conserved, and a randomly transparent interface.
In Section~\ref{SECTpara} we demonstrate that the contribution of
magnon-assisted tunneling to the thermopower of a parallel (P) configuration
is negligible.
In Section~\ref{SECThm} we 
discuss the magnetothermopower, give an order of magnitude estimate
of the size of the effect, and present results for the
magnetothermopower of a junction between two half-metallic ferromagnets.

%
\section{Thermopower of an antiparallel junction}\label{SECTanti}
%

%
\subsection{Description of the model}
%

Our initial aim is to write a balance equation for the current
in terms of occupation numbers of electrons
$n_{L(R)} (\epsilon_{{\bf k}\alpha}) =
[\exp ((\epsilon_{{\bf k}\alpha} - \epsilon_F^{L(R)})/(k_BT_{L(R)})) + 1]^{-1}$
and of magnons
$N_{L(R)} ({\bf q}) =
[\exp (\omega_{\bf q}/(k_BT_{L(R)})) - 1]^{-1}$
on the left (right) hand side of the junction, where
$T_{L(R)}$ is the temperature on the left (right) hand side,
$\epsilon_F^{L} - \epsilon_F^{R} = -eV$,
and $\omega_{\bf q}$ is the energy of a magnon of wavevector ${\bf q}$.
In the following we set $T_L = T$ and $T_R = T - \Delta T$
and we shall speak throughout in terms
of the transfer of electrons with charge $-e$.
The index $\alpha = \{ + , - \}$ takes account of the splitting of
conduction band electrons into majority $\epsilon_{{\bf k}+}$ and
minority $\epsilon_{{\bf k}-}$ subbands,
$\epsilon_{{\bf k}\alpha} =
\epsilon_{\bf k} - \alpha\Delta/2$,
where $\epsilon_{\bf k}$ is the bare electron energy and
$\Delta$ is the spin splitting energy.

In an AP junction, we assume that the majority electrons on the left hand
side of the junction are `spin-up' and the majority electrons on the right
are `spin-down'. The total Hamiltonian of the system is
\begin{eqnarray}
H &=& H_F^L + H_F^R + H_T ,  \label{Htot} \\
H_{T} &=&
\sum_{{\bf k k^{\prime}}\alpha}
\left[ t_{{\bf k},{\bf k^{\prime}}}
c_{{\bf k}\alpha}^{\dag} c_{{\bf k^{\prime}}\overline{\alpha}}
+ t_{{\bf k},{\bf k^{\prime}}}^{*}
c_{{\bf k^{\prime}}\overline{\alpha}}^{\dag} c_{{\bf k}\alpha} \right] ,
\label{ht} 
\end{eqnarray}
where $H_T$ is the tunneling Hamiltonian\cite{coh62,car72,mah81}
that is written in terms
of creation and annihilation Fermi operators $c^{\dag}$ and $c$.
Here $\overline{\alpha} \equiv - \alpha$ and
we assume that spin is conserved when an electron
tunnels across the interface.
The tunneling matrix elements $t_{{\bf k},{\bf k^{\prime}}}$ describe the
transfer of an electron with wavevector ${\bf k}$ on the left
to the state with ${\bf k^{\prime}}$ on the right.
We will consider $t_{{\bf k},{\bf k^{\prime}}}$ to be a symmetric matrix of
the form
\begin{eqnarray}
t_{{\bf k},{\bf k^{\prime}}} &=&
\tilde{t}_{{\bf k}_{||}, {\bf k^{\prime}}_{||}}
\left|\frac{h^2  v_L^z({\bf k})  v_R^z({\bf k^{\prime}})}{L^2}\right|^{1/2} \!,
\label{t}
\end{eqnarray}
where $v_{L,R}^z({\bf k}) =
\partial\epsilon_{L,R}({\bf k})/\partial(\hbar k_z)$
are components of electron velocity perpendicular to the interface and
$\epsilon_{L,R}({\bf k})$
denotes the electron energy dispersion in the electrodes.
In our model for $t$, we neglect its explicit energy dependence.
However, $\tilde{t}_{{\bf k}_{||}, {\bf k^{\prime}}_{||}}$
can describe both clean and diffusive interfaces by taking into account the
conservation of ${\bf k}_{||}$, the component of momentum parallel
to the interface.

The term $H_F^{L(R)}$ is the Hamiltonian of the ferromagnetic electrode
on the left (right) side of the junction in the absence of tunneling. 
We use the so called s-f (s-d) model,\cite{w+w70,nag74} which assumes that
magnetism and electrical conduction are caused by different
groups of electrons that are coupled via an intra-atomic exchange interaction,
although we note that the same results, in the lowest order of
electron-magnon interactions, may be obtained from a model of itinerant
ferromagnets.\cite{e+h73}
The magnetism originates from inner atomic 
shells (e.g., d or f) which have unoccupied electronic orbitals and,
therefore, possess magnetic moments whereas the 
conduction is related to electrons with delocalized wave functions.
Using the Holstein-Primakoff transformation\cite{h+p40} the 
operators of the localized moments in the interaction Hamiltonian
can be expressed via magnon creation and 
annihilation  operators $b^\dag,  b$.
At low temperatures, where the average number of magnons is small
$<b^\dag  b>\ll 2\xi$ ($\xi$ is the spin of the localized moments),
the Hamiltonian of the ferromagnet $H_F^{L(R)}$ can be written as follows
\begin{eqnarray}
H_F^{L(R)} = H_e^{L(R)} + H_m^{L(R)} + H_{em}^{L(R)},
\label{hf} 
\end{eqnarray}
\begin{eqnarray}
H_e^{L(R)} = \sum_{{\bf k}\alpha}\epsilon_{{\bf k}\alpha}
c^\dag_{{\bf k}\alpha}c_{{\bf k}\alpha},
\qquad 
\epsilon_{{\bf k}\alpha} =
\epsilon_{\bf k} - \alpha\Delta/2,
\label{he}  
\end{eqnarray}
\begin{eqnarray}
H_m^{L(R)} = \sum_{{\bf q}} \omega_{{\bf q}} b^{\dag}_{{\bf q}} b_{{\bf q}},
\qquad \omega_{q=0}=\omega_0,
\label{hm}  
\end{eqnarray}
\begin{eqnarray}
H_{em}^{L(R)} = \!
- \frac{\Delta}{\sqrt{2 \xi\cal{N}}} \!
\sum_{{\bf k q}} \!
\left[ c_{{\bf k-q}+}^{\dag} c_{{\bf k}-} b_{{\bf q}}^{\dag}
+ c_{{\bf k}-}^{\dag} c_{{\bf k-q}+} b_{{\bf q}} \right] , \!\!
\label{hem}
\end{eqnarray}
The first term $H_e^{L(R)}$, Eq.~(\ref{he}),
deals with conduction band electrons which are split into majority 
$\epsilon_{{\bf k}+}$ and minority $\epsilon_{{\bf k}-}$ subbands due to
the s-f (s-d) exchange.
The Hamiltonian $H_m^{L(R)}$, Eq.~(\ref{hm}),
describes free magnons with spectrum $\omega_q$ which in the general case
has a gap $\omega_{q=0}=\omega_0$.
The third term $H_{em}^{L(R)}$, Eq.~(\ref{hem}),
is the electron-magnon coupling resulting from the intra-atomic exchange
interaction between the spins of the conduction electrons
and the localized moments.

The calculation is performed using standard second order perturbation
theory.
We write the total Hamiltonian, Eq.~(\ref{Htot}), as $H = H_0 + V$, 
where the perturbation $V = H_{T} + H_{em}^L + H_{em}^R$ is the sum
of the tunneling Hamiltonian and the electron-magnon interactions in the
electrodes.
First order terms provide an elastic contribution
to the current that do not involve any change of the spin orientation of the
itinerant electrons, whilst second order terms account for inelastic,
magnon-assisted processes.

%
\subsection{Elastic contribution to the current}
%
The first order contribution to the current,
in the antiparallel configuration, arises from elastic tunneling
without any spin flip between a majority conduction electron state on one
side of the junction and a minority state on the other.
Consider for example an initial state consisting of an additional
majority spin up electron on the left with wavevector ${\bf k_L}$
and energy $\epsilon_{{\bf k_L}+}$.
This electron can tunnel, with matrix element $t_{LR}^{*}$,
into a minority spin up state on the right with
wavevector ${\bf k_R}$ and energy $\epsilon_{{\bf k_R}-}$.
In addition there is a second process which is a transition between the same
two states, but in the reverse order, giving a contribution to the current
with an opposite sign.
Together, the two processes give a balance equation for the first order
contribution to the current between the majority band on the left and the
minority on the right.
In addition, there are two first order processes that result in transitions
between the minority band on the left and the majority on the right.
Overall, the first order contribution to the current is $I_{AP}^{(1)}$ where
\begin{eqnarray}
I_{AP}^{(1)} &=& - 4 \pi^2 \frac{e}{h}
\int_{-\infty}^{+\infty} \!\! d\epsilon
\sum_{{\bf k_L}{\bf k_R}} \sum_{\alpha = \{ \pm \}}
\left| t_{{\bf k_L},{\bf k_R}} \right|^2
\delta (\epsilon - \epsilon_{{\bf k_L}\alpha} ) 
\nonumber \\
&& \times
\, \delta (\epsilon - eV - \epsilon_{{\bf k_R}\overline{\alpha}} )
\Big\{
n_L (\epsilon_{{\bf k_L}\alpha})
\left[ 1 - n_R (\epsilon_{{\bf k_R}\overline{\alpha}}) \right] -
\nonumber \\
&& \qquad \qquad
- \left[ 1 - n_L (\epsilon_{{\bf k_L}\alpha}) \right]
n_R (\epsilon_{{\bf k_R}\overline{\alpha}})
\Big\} ,
\label{i1a}
\end{eqnarray}
and $\overline{\alpha} \equiv - \alpha$.
Neglecting terms that contain the small parameter $k_B T/\epsilon_F$,
the current may be written as
\begin{eqnarray}
I_{AP}^{(1)} \approx \frac{e^2}{h} V
\left( {\cal T}_{+-} + {\cal T}_{-+} \right) .
\label{i1ap}
\end{eqnarray}
For convenience we have grouped all the information about the nature of the
interface into a parameter ${\cal T}_{\alpha \alpha^{\prime}}$,
\begin{eqnarray}
{\cal T}_{\alpha \alpha^{\prime}}
\approx
4 \pi^2 \sum_{{\bf k_L}{\bf k_R}}
\left| t_{{\bf k_L},{\bf k_R}} \right|^2
\delta (\epsilon - \epsilon_{\bf k_{L\alpha}} )
\delta (\epsilon - \epsilon_{\bf k_{R\alpha^{\prime}}} ) ,
\label{Tdef}
\end{eqnarray}
that is equivalent to the sum over all scattering channels,
between spin states $\alpha$ on the left and $\alpha^{\prime}$ on the right,
of the transmission eigenvalues usually introduced
in the Landauer formula,\cite{lan70,f+l81,bee97}
although we restrict ourselves to the tunneling regime in this paper.
Later we will employ models of two types of interface explicitly:
a uniformly transparent interface where the component of momentum
parallel to the interface
is conserved, and a randomly transparent interface.
%
\subsection{Magnon-assisted contribution to the current}
%
Below we describe processes which contribute to magnon-assisted tunneling.
For convenience, we divide them into two groups that we
label as `electron' and `hole' processes.
In `electron' processes, an increase in the
number of magnons in one electrode is achieved by accepting
electrons from the other electrode whereas, in `hole' processes,
an increase in the number of magnons in one electrode is achieved by
injecting electrons into the other electrode.

%
\begin{figure}
\epsfxsize=\hsize
\epsffile{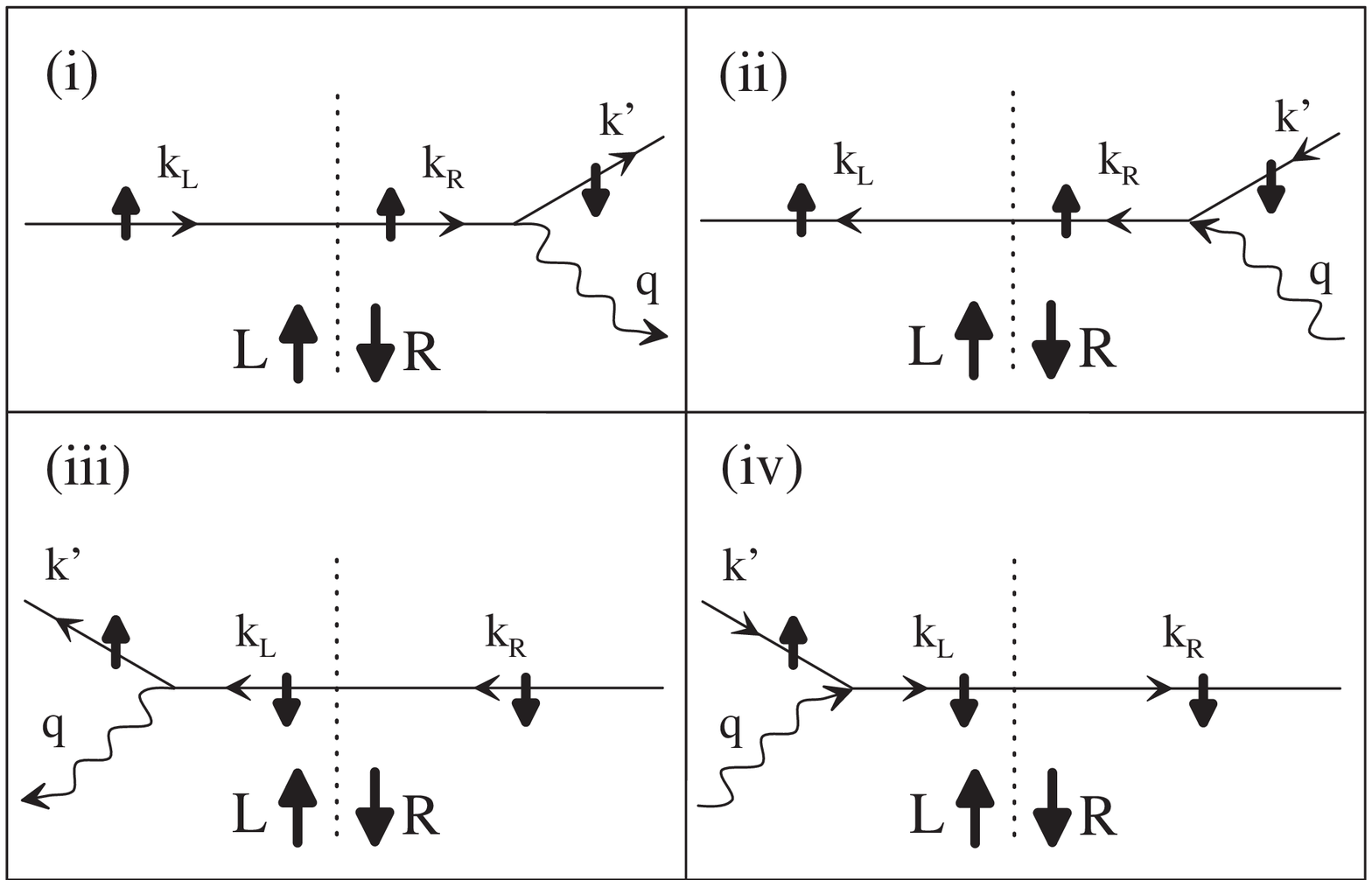}
\refstepcounter{figure}
\label{figure:2}

{\setlength{\baselineskip}{10pt} FIG.\ 2.
Schematic of four `electron' type processes,
across a junction in the antiparallel configuration,
that involve transitions from majority initial to majority final states
via a virtual intermediate state in the minority band:
(i) and (iii) involve magnon emission on the right and left hand sides,
respectively, whereas (ii) and (iv) involve magnon absorption
on the right and left.
}
\end{figure}
%

The `electron' processes are shown schematically in Figure~2.
The straight lines show the direction of electron transfer,
whereas the wavy lines denote the emission or absorption of magnons.
The processes are drawn using the rule, appropriate for ferromagnetic
electron-magnon exchange, that an electron in a minority state
scatters into a majority state by emitting a magnon.
The `electron' processes in Figure~2 involve transitions
from majority initial states to majority final states via
an intermediate, virtual state in the minority band.
For example, process (i), which is the same as the process
shown in more detail in Figure~1, begins with a
spin-up majority electron on the left with wavevector ${\bf k_L}$
and energy $\epsilon_{{\bf k_L}+}$.
Then, this electron tunnels across the barrier (without spin flip) to occupy
a virtual, intermediate state with wavevector ${\bf k_R}$ in the spin up
minority band on the right as depicted in the right part of Figure~1(a)
with energy $\epsilon_{{\bf k_R}-}$.
The energy difference between the states is
$\epsilon_{\bf k_R} - \epsilon_{\bf k_L} + \Delta$
so that the matrix element for the transition contains an energy in the
denominator related to the inverse lifetime of the electron in the virtual
state.
For $k_{B}T,eV \ll \Delta$, when both initial and final electron states should
be taken close to the Fermi level, only long wavelength magnons can be emitted,
so that the energy deficit in the virtual states can be approximated as
$\epsilon_{\bf k_R} - \epsilon_{\bf k_L} + \Delta \approx \Delta$.
As noticed in Refs.~\onlinecite{w+w70,mcc01,tka02},
this cancels out the large exchange parameter since the same electron-core
spin exchange constant appears both in the splitting between minority
and majority bands and in the electron-magnon coupling.

The second part of the transition is sketched in Figure~1(b) where
the electron in the virtual minority spin up state incorporates itself into
a state in the majority spin down band on the right by emitting a magnon,
which is shown as a flip of one of the localized moments.
The wavevector of the electron in the final state is ${\bf k^{\prime}}$ and
the total energy of the final (many body) state is
$\epsilon_{{\bf k^{\prime}}+} - \omega_{\bf q}$.
Similiar considerations enable us to write down the contribution to the current
from all the `electron' processes in Figure~2.
We group the processes into pairs which involve transitions between
the same series of states, but
in the opposite time order so that they give a current with different signs,
hence their sum gives a balance equation.
The contributions to the current of the processes 
(i) and (ii), labelled as $I_{AP}^{\rm (i,ii)}$, are given by
\begin{eqnarray}
I_{AP}^{\rm (i,ii)} &=& - 4 \pi^2 \frac{e}{h}
\int_{-\infty}^{+\infty} \!\! d\epsilon
\sum_{{\bf k_L}{\bf k^{\prime}}{\bf q}}
\frac{\left| t_{{\bf k_L},{\bf k^{\prime}}} \right|^2}{2\xi{\cal N}}
\nonumber \\
&& \times
\,\delta (\epsilon - \epsilon_{{\bf k_L}+}) 
\,\delta (\epsilon - eV - \epsilon_{{\bf k^{\prime}}+} - \omega_{\bf q})
\nonumber \\
&& \times
\Big\{
n_L (\epsilon_{{\bf k_L}+})
\left[ 1 - n_R (\epsilon_{{\bf k^{\prime}}+}) \right]
\left[ 1 + N_R ({\bf q}) \right] -
\nonumber \\
&& - \left[ 1 - n_L (\epsilon_{{\bf k_L}+})\right]
n_R (\epsilon_{{\bf k^{\prime}}+}) N_R ({\bf q})
\Big\} ,
\label{iab}
\end{eqnarray}
where ${\bf q} = {\bf k_R} - {\bf k^{\prime}}$.
The processes (iii) and (iv) are similiar to processes
(i) and (ii), respectively, except that electrons interact with magnons
in the left electrode:
\begin{eqnarray}
I_{AP}^{\rm (iii,iv)} &=& - 4 \pi^2 \frac{e}{h}
\int_{-\infty}^{+\infty} \!\! d\epsilon
\sum_{{\bf k^{\prime}}{\bf k_R}{\bf q}}
\frac{\left| t_{{\bf k^{\prime}},{\bf k_R}} \right|^2}{2\xi{\cal N}}
\nonumber \\
&& \times \,\delta (\epsilon - eV - \epsilon_{{\bf k_R}+} )
\,\delta (\epsilon - \epsilon_{{\bf k^{\prime}}+} - \omega_{\bf q})
\nonumber \\
&& \times \Big\{
- n_R (\epsilon_{{\bf k_R}+})
\left[ 1 - n_L (\epsilon_{{\bf k^{\prime}}+}) \right]
\left[ 1 + N_L({\bf q}) \right] +
\nonumber \\
&& +
\left[ 1 - n_R (\epsilon_{{\bf k_R}+}) \right]
n_L (\epsilon_{{\bf k^{\prime}}+}) N_L ({\bf q})
\Big\},
\label{icd}
\end{eqnarray}
where ${\bf q} = {\bf k_L} - {\bf k^{\prime}}$.
Energies on the right are shifted by $eV$ to take account of the
voltage difference across the junction.

An example of a `hole' process is shown in detail in Figure~3 (a) and (b),
and Figure~3(c) shows the same process, (v), plus other  
`hole' processes (vi), (vii), (viii).
The `hole' processes
involve transitions from minority initial to minority final states
via a virtual intermediate state in the majority band.
In contrast to the `electron' processes, an increase in the number of
magnons in one electrode is achieved by injecting electrons into the
other electrode.
As an example we describe in detail the calculation of the matrix element
for process (v) shown in Figure~3 (a) and (b).
The initial state has an additional spin down minority
electron near the Fermi level on the left (left part of Figure~3(a))
with wavevector ${\bf k_L}$.
The first step in the transition is the creation of an empty
state below the Fermi level in the spin down majority band on the right
with wavevector ${\bf k_R}$ by the absorption of a magnon,
wavevector ${\bf q}$, to elevate
a spin down majority electron up to a
spin up minority state above the Fermi level on the right
with wavevector ${\bf k^{\prime}}$ (right part of Figure~3(a)).
The second part of the transition is sketched in Figure~3(b) where
the spin down minority electron on the left tunnels across the barrier
(without spin flip) to occupy the empty spin down majority state
on the right.
The contributions to the current from
processes (v) and (vi), $I_{AP}^{\rm (v,vi)}$,
and from processes (vii) and (viii), $I_{AP}^{\rm (vii,viii)}$, are

%
\begin{figure}
\hspace{0.05\hsize}
\epsfxsize=0.8\hsize
\epsffile{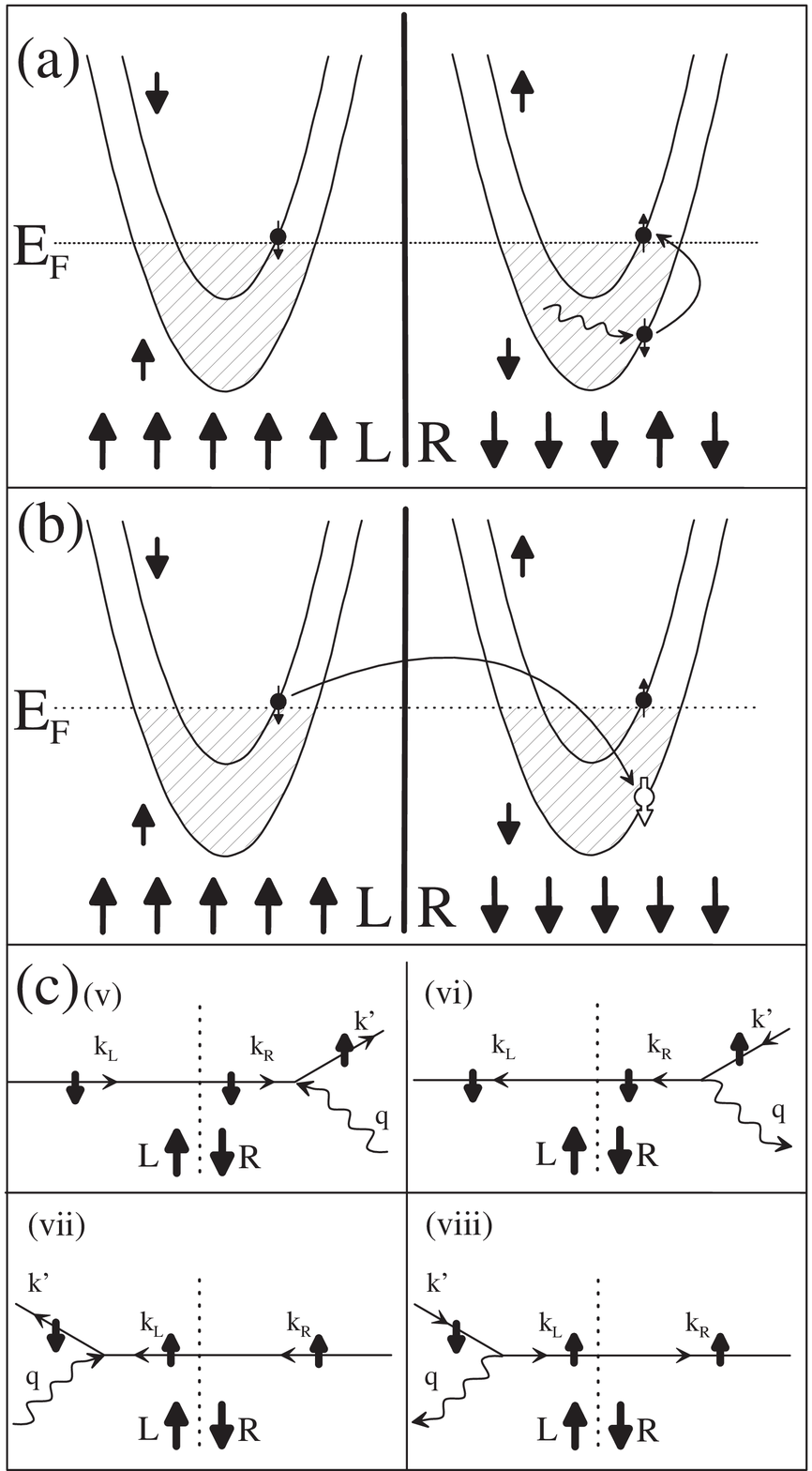}
\refstepcounter{figure}
\label{figure:3}

{\setlength{\baselineskip}{10pt} FIG.\ 3.
Schematic of `hole' type processes of
magnon-assisted tunneling from an initial minority state to
a final minority state via an intermediate majority state.
(a) A typical process begins with the absorption of a magnon (wavy line)
to elevate a spin-down majority electron below the Fermi level
$E_F$ on the right up to a spin-up minority state above $E_F$,
creating an empty state below $E_F$ in the spin-down majority band.
(b) A spin-down minority electron on the left tunnels across the barrier
(without spin flip) to occupy the empty spin down majority state
on the right.
(c) The same process (v) plus remaining `hole' processes:
(v) and (vii) involve magnon absorption on the right and left hand sides,
respectively, whereas (vi) and (viii) involve magnon emission
on the right and left.}
\end{figure}
%

%
\begin{eqnarray}
I_{AP}^{\rm (v,vi)} &=& - 4 \pi^2 \frac{e}{h}
\int_{-\infty}^{+\infty} \!\! d\epsilon
\sum_{{\bf k_L}{\bf k^{\prime}}{\bf q}}
\frac{\left| t_{{\bf k_L},{\bf k^{\prime}}} \right|^2}{2\xi{\cal N}}
\nonumber \\
&& \times
\,\delta (\epsilon - \epsilon_{{\bf k_L}-} ) 
\,\delta (\epsilon - eV - \epsilon_{{\bf k^{\prime}}-} + \omega_{\bf q})
\nonumber \\
&& \times
\Big\{
n_L (\epsilon_{{\bf k_L}-})
\left[ 1 - n_R (\epsilon_{{\bf k^{\prime}}-}) \right]
N_R ({\bf q}) -
\nonumber \\
&& - \left[ 1 - n_L (\epsilon_{{\bf k_L}-})\right]
n_R (\epsilon_{{\bf k^{\prime}}-})
\left[ 1 + N_R ({\bf q}) \right]
\Big\} .
\label{ief}
\end{eqnarray}
\begin{eqnarray}
I_{AP}^{\rm (vii,viii)} &=& - 4 \pi^2 \frac{e}{h}
\int_{-\infty}^{+\infty} \!\! d\epsilon
\sum_{{\bf k^{\prime}}{\bf k_R}{\bf q}}
\frac{\left| t_{{\bf k^{\prime}},{\bf k_R}} \right|^2}{2\xi{\cal N}}
\nonumber \\
&& \!\!\! \times \,\delta (\epsilon - eV - \epsilon_{{\bf k_R}-} )
\,\delta (\epsilon - \epsilon_{{\bf k^{\prime}}-} + \omega_{\bf q})
\nonumber \\
&& \!\!\! \times \Big\{
- n_R (\epsilon_{{\bf k_R}-})
\left[ 1 - n_L (\epsilon_{{\bf k^{\prime}}-}) \right]
N_L ({\bf q}) +
\nonumber \\
&& \!\! +
\left[ 1 - n_R (\epsilon_{{\bf k_R}-}) \right]
n_L (\epsilon_{{\bf k^{\prime}}-})
\left[ 1 + N_L({\bf q}) \right]
\Big\},
\label{igh}
\end{eqnarray}

To make our analysis transparent,
we rewrite the magnon-assisted current as the sum of two parts,
\begin{eqnarray}
I_{AP} &=& I_{AP}^{\rm (i,ii)} + I_{AP}^{\rm (iii,iv)}
+ I_{AP}^{\rm (v,vi)} + I_{AP}^{\rm (vii,viii)} \nonumber \\
&=& I_{AP}^{{\rm spont}} + I_{AP}^{{\rm stim}}, \nonumber
\end{eqnarray}
the first of which, labelled $I_{AP}^{{\rm spont}}$,
does not contain any magnon occupation numbers and represents spontaneous
emission processes,
\begin{eqnarray}
I_{AP}^{{\rm spont}} &=& - \frac{e}{h}
\frac{\left( {\cal T}_{++} + {\cal T}_{--} \right)}{2\xi{\cal N}}
\int_{-\infty}^{+\infty} \!\! d\epsilon
\int_{0}^{\infty} \!\! d\omega \, \Omega (\omega ) \nonumber \\ 
&& \times \Big\{
n_L(\epsilon ) \left[ 1 - n_R (\epsilon - eV - \omega ) \right]
\nonumber \\
&& \quad
- \left[ 1 - n_L (\epsilon - \omega ) \right] n_R(\epsilon - eV) 
\Big\} ,
\label{spont}
\end{eqnarray}
where $\Omega (\omega ) = \sum_{\bf q} \delta (\omega - \omega_{{\bf q}})$
is the magnon density of states that we assume to be the same on both
sides of the junction.
Since our main aim is to demonstrate the existence of an effect, we choose
the simple example of a bulk, three-dimensional magnon density of states.
We assume a quadratic magnon dispersion, $\omega_q = Dq^2$,
and apply the Debye approximation with a maximum magnon
energy $\omega_D = D (6 \pi^2 /v)^{2/3}$
where $v$ is the volume of a unit cell.
This enables us to express the magnon density of states as
$\Omega (\omega ) = (3/2) {\cal N} \omega^{1/2} / \omega_D^{3/2}$.
The term $I_{AP}^{{\rm spont}}$ is only non-zero for finite voltage.
We calculate it in two different limits, small temperature difference
$\Delta T \ll T$ and large temperature difference
$\Delta T \!=\!  T, T_R = 0$.
\begin{eqnarray}
I_{AP}^{{\rm spont}} &=& \frac{e^2}{h} \frac{V}{\xi}
\left( {\cal T}_{++} + {\cal T}_{--} \right)
\left( \frac{k_BT_L}{\omega_D} \right)^{3/2}
\nonumber \\
&& \, \times
\frac{3}{4} \Gamma (\textstyle\frac{3}{2}) \zeta (\textstyle\frac{3}{2})
\times \left\{
\begin{array}{l@{\quad  \quad}l}
1 \, ,
& \Delta T \ll T \\
2 - \!\sqrt{2} \, ,
& \Delta T \!=\! T
\end{array}
\right.
,
\label{ispont}
\end{eqnarray}
where $\Gamma (x)$ is the gamma function and
$\zeta (x)$ is Riemann's zeta function.\cite{g+r}

The second term, labelled $I_{AP}^{{\rm stim}}$, contains all the magnon
occupation numbers and it represents absorption and
stimulated emission processes,
\begin{eqnarray}
&& I_{AP}^{{\rm stim}} = - \frac{e}{h}
\frac{1}{2\xi{\cal N}}
\int_{-\infty}^{+\infty} \!\! d\epsilon
\int_{0}^{\infty} \!\! d\omega \, \Omega (\omega ) 
\nonumber \\
&&  \times
\Big\{
\left[ n_L (\epsilon ) - n_R (\epsilon - eV + \omega ) \right] \;\!\!
\left[ {\cal T}_{++} N_L(\omega ) + {\cal T}_{--} N_R(\omega ) \right]
 \nonumber \\
&& \, +
\left[ n_L (\epsilon ) - n_R (\epsilon - eV - \omega ) \right] \;\!\!
\left[ {\cal T}_{++} N_R(\omega ) + {\cal T}_{--} N_L(\omega ) \right]
\Big\} .
\nonumber \\
\label{stim}
\end{eqnarray}
It vanishes in the limit of zero temperature on both sides of the
junction, but is non-zero for zero bias voltage in the presence of
a temperature difference.
$I_{AP}^{{\rm stim}}$ may be written explicitly for arbitrary bias
voltage and temperatures,
\begin{eqnarray}
&& I_{AP}^{{\rm stim}} = \frac{e}{h}
\frac{3}{4\xi\omega_D^{3/2}} \times
\nonumber \\
&& \times
\Big\{
eV \:\!\!\left( {\cal T}_{++} + {\cal T}_{--} \right)
\;\!\! \Gamma (\textstyle\frac{3}{2})
\zeta (\textstyle\frac{3}{2}) \!
\left[ (k_BT_L)^{3/2}\! + (k_BT_R)^{3/2} \right] -
\nonumber \\
&& - \left( {\cal T}_{++} - {\cal T}_{--} \right)
\;\!\! \Gamma (\textstyle\frac{5}{2}) \zeta (\textstyle\frac{5}{2}) \!
\left[ (k_BT_L)^{5/2}\! - \!(k_BT_R)^{5/2} \right] \!
\Big\} .
\label{stim3}
\end{eqnarray}
%
%
\subsection{Calculation of the thermopower}
%
The thermopower $S_{AP}$ is determined by setting the total current to zero,
$I_{AP}^{(1)} + I_{AP}^{{\rm spont}} + I_{AP}^{{\rm stim}} = 0$,
and finding the voltage $V$ induced by the temperature
difference $\Delta T$, $S_{AP} = - V/\Delta T$.
In general we find
\begin{eqnarray}
S_{AP} = - \frac{k_B}{e}
\frac{C \left( {\cal T}_{++} - {\cal T}_{--} \right)  \delta m(T)}
{\left[ {\cal T}_{+-} + {\cal T}_{-+}
+ B \:\!\!
\left( {\cal T}_{++} + {\cal T}_{--} \right) \:\!\! \delta m(T) \right]} .
\label{Sall}
\end{eqnarray}
This is the main result of the paper, describing junctions between
ferromagnets of arbitrary polarization strength ranging from
weak ferromagnets to half-metals.
The factors $C$ and $B$ are dependent on the ratio
$\Delta T/T$. We evaluate them in the limits of small, $\Delta T \ll T$,
and large, $\Delta T = T_L = T$, $T_R = 0$, temperature difference:
\begin{eqnarray}
C = \frac{\zeta (\textstyle\frac{5}{2})}{\zeta (\textstyle\frac{3}{2})}
\times
\left\{
\begin{array}{l@{\quad  \quad}l}
15/8 \, ,
& \Delta T \ll T \\
3/4 \, ,
& \Delta T \!=\! T 
\end{array}
\right. ,
\label{Cfactor}
\end{eqnarray}
\begin{eqnarray}
B = \left\{
\begin{array}{l@{\quad  \quad}l}
3/2 \, ,
& \Delta T \ll T \\
(3-\sqrt{2})/2 \, ,
& \Delta T \!=\! T 
\end{array}
\right. .
\label{Bfactor}
\end{eqnarray}
The function $\delta m(T)$ in Eq.~(\ref{Sall}) is the change in the
magnetization due to thermal magnons at temperature $T$
(Bloch's $T^{3/2}$ law),\cite{kittel}
\begin{eqnarray}
\delta m(T)
&=& \frac{1}{\xi\cal{N}} \int_{0}^{\infty} d\omega \,
\Omega (\omega ) N_L (\omega )
\nonumber \\
&=&
\frac{3}{2\xi} \:\!
\Gamma (\textstyle\frac{3}{2}) \zeta (\textstyle\frac{3}{2})
\!\!\:
\left( \frac{k_BT}{\omega_D} \right)^{3/2} .
\label{mt}
\end{eqnarray}
The thermopower is finite because the current
$I_{AP}^{{\rm stim}}$ contains a term (last line in Eq.~(\ref{stim3}))
that depends on the temperature difference.
It arises from the difference in the thermal distribution of magnons
$N_L(\omega ) - N_R(\omega )$ and the process of thermal equilibration
between two baths of magnons held at different temperatures,
which is mediated by electrons, results in a current.
The origin of the factor ${\cal T}_{++} - {\cal T}_{--}$ in the numerator 
of Eq.~(\ref{Sall}) can be understood in the following way.
In the `electron' processes in Figure~2, which contribute to ${\cal T}_{++}$, 
an increase in the number of magnons in one electrode is achieved by accepting
electrons from the other electrode.
On the other hand, in the the `hole' processes in Figure~3,
which contribute to ${\cal T}_{--}$, an increase in the
number of magnons in one electrode is achieved by injecting
electrons into the other electrode.
Hence the contributions of ${\cal T}_{++}$ and ${\cal T}_{--}$
proportional to $\Delta T$ in the current Eq.~(\ref{stim3}) appear
with opposite signs.

The sign of the thermopower, Eq.~(\ref{Sall}), is
specified for electron (charge $-e$) transfer processes and under
the assumption that the exchange between conduction band and core
electrons has a ferromagnetic sign.
For antiferromagnetic exchange, the overall sign of the
thermopower would be opposite.
For example, processes (i) and (ii) in Fig.~2, would
involve magnons on the opposite side of the junction,
hence the current $I_{AP}^{{\rm (i,ii)}}$ would be determined by
magnon occupation numbers $N_{{\rm L}}({\bf q})$.
Note also that we considered a bulk, three-dimensional 
magnon density of states,
but in general the magnitude and sign of the thermopower will depend
on the magnon spectrum.
%
\subsubsection{Uniformly transparent interface}
%
We model different types of interface by introducing a dependence of
the tunneling matrix elements
$t_{{\bf k_L}, {\bf k_R}}$ on the wavevectors
${\bf k_L} = ( {\bf k_{L}^{||}} , k_{L}^{z} )$ and
${\bf k_R} = ( {\bf k_{R}^{||}} , k_{R}^{z} )$, where
${\bf k_{L(R)}^{||}}$ is the component parallel to the interface and
$k_{L(R)}^{z}$ is the perpendicular component. 
For a uniformly transparent interface of area $A$ such that the parallel
component of momentum is conserved, we set the dimensionless tunneling
factor, Eq.~(\ref{t}), equal to
\begin{eqnarray}
| \tilde{t}_{{\bf k_{L}^{||}},{\bf k_{R}^{||}}} |^2 = \left| t \right|^2
\delta_{{\bf k_{L}^{||}},{\bf k_{R}^{||}}} \, ,
\end{eqnarray}
so that
\begin{eqnarray}
{\cal T}_{\alpha \alpha^{\prime}}
\approx 4 \pi^2 \left| t \right|^2
\frac{A}{h^2} \, {\rm min} \{ \Pi_{\alpha} , \Pi_{\alpha^{\prime}} \}
\end{eqnarray}
where $t$ represents the transparency of the interface
and $\Pi_{\alpha}$ is the area of the maximal cross-section
of the Fermi surface of spins $\alpha$, $\Pi_{+} > \Pi_{-} > 0$.
Then ${\cal T}_{+-} = {\cal T}_{-+} = {\cal T}_{--}
= 4 \pi^2 \left| t \right|^2 (A\Pi_{-}/h^2)$
and
${\cal T}_{++} = 4 \pi^2 \left| t \right|^2 (A\Pi_{+}/h^2)$.
In the regime
$1 \gg \!(1 + \Pi_{+}/\Pi_{-}) \,\delta m(T)$,
the thermopower Eq.~(\ref{Sall}) simplifies as
\begin{eqnarray}
S_{AP} &=& - C \frac{k_B}{e}
\frac{\left( \Pi_{+} - \Pi_{-} \right)}{2\Pi_{-}}
\, \delta m(T)  .
\end{eqnarray}
%
%
\subsubsection{Diffusive tunnel barrier}
%
We use a model\cite{tka02} for describing a strongly nonuniform
interface which is transparent in a finite number of points only,
randomly distributed over an area A.
Each transparent point is treated as a defect which causes electron
scattering in the plane parallel to the interface and
the tunneling matrix element is a matrix element of the total scattering
potential determined with the use of plane waves,
\begin{eqnarray}
\tilde{t}_{{\bf k_{L}^{||}},{\bf k_{R}^{||}}} = \frac{a}{A} \sum_j t_j
\exp \left[
i \hbar^{-1} ({\bf k_L^{||}} - {\bf k_R^{||}}).{\bf r_j}
\right] ,
\end{eqnarray}
where ${\bf r_j}$ is the position of the $j$th contact
with area $a \sim \lambda_F^2$.
A product of two tunneling matrix elements, averaged with respect
to the position of each defect, will be large only
if the total phase shift is zero which corresponds to scattering
from the same defect,
\begin{eqnarray}
\langle |\tilde{t}_{{\bf k_{L}^{||}},{\bf k_{R}^{||}}}|^2 \rangle
= \left( \frac{a}{A} \right)^2 |t|^2 ,
\end{eqnarray}
where $a/A$ accounts for a reduced effective area and $|t|^2 = \sum_j |t_j|^2$
is an effective transparency.
This means that
\begin{eqnarray}
{\cal T}_{\alpha \alpha^{\prime}}
\approx 4 \pi^2
\left| t \right|^2
\left( \frac{a\Pi_{\alpha}}{h^2} \right)
\left( \frac{a\Pi_{\alpha^{\prime}}}{h^2} \right) .
\end{eqnarray}
We assume that the densities of states in the spin band $\alpha$ are
equal on both sides of the junction so that ${\cal T}_{+-} = {\cal T}_{-+}$.
In the regime
$1 \gg \!(\Pi_{+}/\Pi_{-} + \Pi_{-}/\Pi_{+}) \,\delta m(T)$,
the thermopower Eq.~(\ref{Sall}) simplifies as
\begin{eqnarray}
S_{AP} &=& - C \frac{k_B}{e}
\frac{\left( \Pi_{+}^2 - \Pi_{-}^2 \right)}{2 \Pi_{+}\Pi_{-}}
\, \delta m(T)  .
\end{eqnarray}
%
%
%
\section{Thermopower of a parallel junction}\label{SECTpara}
%
For parallel orientation of the magnetic polarizations of the
ferromagnets, we find that the contribution of magnon-assisted
tunneling to the thermopower is zero
(upto the lowest order in the electron-magnon interaction).
We consider the majority electrons on both sides of the
junction to be spin up and the minority electrons to be spin down,
and the technical details of the calculation of the current are similiar to
those described previously for the antiparallel orientation.
There is a first order, elastic contribution $I_{P}^{(1)}$,
\begin{eqnarray}
I_{P}^{(1)} \approx \frac{e^2}{h} V
\left( {\cal T}_{++} + {\cal T}_{--} \right)
+ {\cal O} \left( \frac{e}{h}
\frac{k_B^2T\Delta T}{\epsilon_F} \right) ,
\label{i1p}
\end{eqnarray}
involving tunneling between majority states on the left and right,
${\cal T}_{++}$, and tunneling between minority states, ${\cal T}_{--}$,
without any spin flip processes.
The first term in Eq.~(\ref{i1p}) corresponds to a large current response
to finite voltage whereas the second term arises from the energy
dependence of the electronic density of states near the Fermi energy.

%
\begin{figure}
\epsfxsize=\hsize
\epsffile{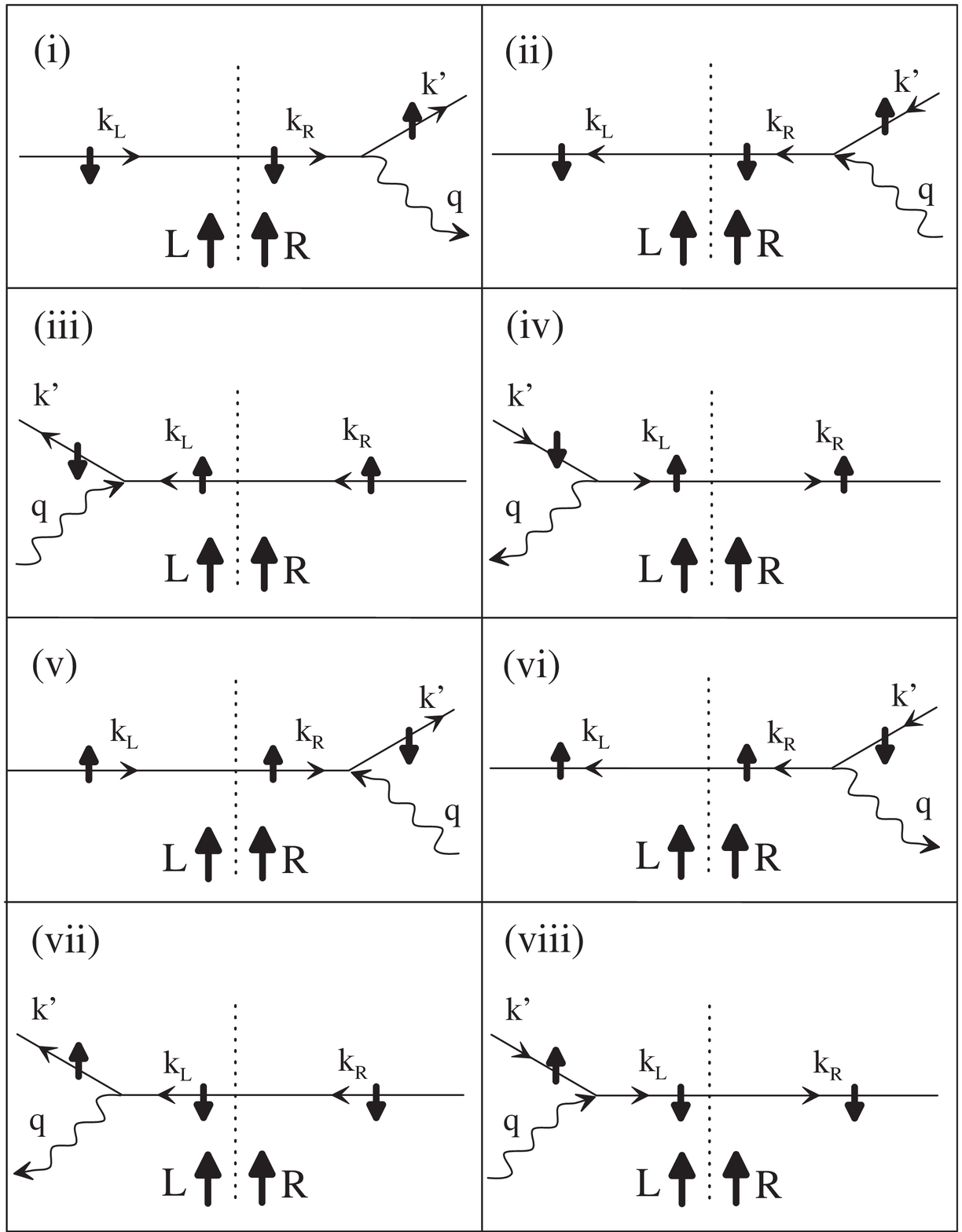}
\refstepcounter{figure}
\label{figure:4}

{\setlength{\baselineskip}{10pt} FIG.\ 4.
Schematic of magnon-assisted tunneling across a junction with ferromagnetic
electrodes in the parallel configuration.
Eight processes which, to lowest order in the electron-magnon interaction,
contribute to magnon-assisted tunneling.
(i) - (iv) involve transitions between minority states on
the left and majority states on the right
via a virtual intermediate state.
(v) - (viii) involve transitions between majority states on
the left and minority states on the right
via a virtual intermediate state.
}
\end{figure}
%

To the lowest order in the electron-magnon interaction there are eight
magnon-assisted tunneling processes which are shown schematically
in Figure~4.
The top four processes, (i)-(iv), involve transitions between minority
states on the left and majority states on the right,
whereas the lower four processes, (v)-(viii), involve transitions between
majority states on the left and minority states on the right.
The overall contribution to the thermopower is zero because
the stimulated emission part of the current does not depend on
the temperature difference across the junction,
\begin{eqnarray}
&& I_{P}^{{\rm stim}} = \frac{e}{h}
\frac{3}{4\xi\omega_D^{3/2}} \times
\nonumber \\
&& \times
\Big\{
eV \;\!\! \left( {\cal T}_{+-} + {\cal T}_{-+} \right)
\;\!\! \Gamma (\textstyle\frac{3}{2})
\zeta (\textstyle\frac{3}{2}) \!
\left[ (k_BT_L)^{3/2}\! + (k_BT_R)^{3/2} \right] -
\nonumber \\
&& - \left( {\cal T}_{+-} - {\cal T}_{-+} \right)
\;\!\! \Gamma (\textstyle\frac{5}{2}) \zeta (\textstyle\frac{5}{2}) \!
\left[ (k_BT_L)^{5/2}\! + \!(k_BT_R)^{5/2} \right] \!
\Big\} .
\label{Pstim}
\end{eqnarray}
This can be understood by examining the processes in Figure~4.
The top four processes (i)-(iv), which have minority states on the left
and majority on the right, all produce terms proportional
to ${\cal T}_{-+}$.
Two of them, (i) and (ii), are `electron' type processes in which an
increase (decrease) in the number of magnons on the right is achieved
by accepting (injecting) electrons from (into) the left,
but the other two, (iii) and (iv), are `hole' type processes in which an
increase (decrease) in the number of magnons on the left is achieved by
injecting (accepting) electrons into (from) the right.
Therefore the term proportional to ${\cal T}_{-+}$ in the last line of
$I_{P}^{{\rm stim}}$, Eq.~(\ref{Pstim}), does not depend on
the temperature difference but a sum
$(k_BT_L)^{5/2} + (k_BT_R)^{5/2}$.
The same is true for the lower four processes in Figure~4, (v) - (viii),
that produce terms proportional to ${\cal T}_{+-}$.

%
\section{Conclusion}\label{SECThm}
%
As shown above, the thermopower of a tunnel F-F junction
in the parallel configuration, $S_P \sim k_B^2 T/(e\epsilon_F)$,
is smaller than the contribution of magnon-assisted transport to the
thermopower $S_{AP}$ in the antiparallel configuration, Eq.~(\ref{Sintro}).
As the relative polarizations of ferromagnetic layers can be manipulated by
an external magnetic field,
the large difference $\Delta S = S_{AP} - S_P$
results in a magnetothermopower effect.
As a rough estimate, we take $\epsilon_F = 5$eV and
$\delta m = 7.5 \!\times \! 10^{-6} \, T^{3/2}$
(for a ferromagnet such as Ni, Ref.~\onlinecite{kittel})
to give $S_{AP} \sim - 3 {\rm\mu \;\!\! V\,K^{-1}}$
and $S_P \sim 0.5 {\rm\mu \;\!\! V\,K^{-1}}$
at $T = 300$K.

As an extreme example, we predict a giant magnetothermopower for a junction
between two half-metallic ferromagnets.
In a half-metal the
splitting $\Delta$ between the majority and minority conduction bands
is greater than $\epsilon_F$ measured from the bottom of the majority band
so that only majority carriers are present at the
Fermi energy.
In this case ${\cal T}_{+-} = {\cal T}_{-+} = {\cal T}_{--} = 0$
in Eq.~(\ref{Sall}) and, in the linear regime $\Delta T \ll T_L$,
\begin{eqnarray}
S_{AP} = - 0.64 \frac{k_B}{e}  \, ; \qquad 
S_{P} \sim \frac{k_B^2 T}{e\epsilon_F} .
\label{shalf}
\end{eqnarray}
This result is independent of temperature and of the specific half-metallic
material, and it represents a giant magnetothermopower effect 
$\Delta S \approx S_{AP} \approx - 55 {\rm\mu \;\!\! V\,K^{-1}}$.

A strong polarization dependence of the thermopower,
$\Delta S\approx S_{AP}$, enables one to separate the interface contribution
to the thermopower from effects arising from a finite temperature gradient
in the reservoirs. We assume in our analysis that phonon-mediated heat
conduction is much lower than the electronic one, and that a fast temperature
equilibration inside the ferromagnetic metal makes a finite temperature drop
across the tunnel barrier possible.
Furthermore, the predicted interface magnetothermopower
will be most pronounced in a geometry where the bottleneck for electron
transport is also the bottleneck for thermal transport: in a small area
mesoscopic junction, ideally, in a suspended STM-type geometry.

 
The authors thank G.~Tkachov and A.~Geim for discussions.
This work was supported by EPSRC, the Royal Society, and the EU High Field
Infrastructure Cooperative Network.


\end{multicols}
\end{document}